\documentclass[twocolumn,journal]{IEEEtran}

\usepackage[T1]{fontenc}
\usepackage[cmex10]{amsmath}
\usepackage{amssymb}
\usepackage{amsthm}
\usepackage{acronym}
\usepackage{threeparttable}
\usepackage{booktabs}
\usepackage[pdftex]{graphicx}
\usepackage{cite}
\usepackage{color}

\usepackage{comment}

\usepackage{subcaption}
\usepackage{float}
\usepackage{makecell}
\usepackage[font=small]{caption}

\usepackage[unicode=true,
bookmarks=true,bookmarksnumbered=true,bookmarksopen=true,bookmarksopenlevel=1,
 breaklinks=false,pdfborder={0 0 0},pdfborderstyle={},backref=false,colorlinks=false]
 {hyperref}
\hypersetup{pdftitle={Your Title},
 pdfauthor={Your Name},
 pdfpagelayout=OneColumn, pdfnewwindow=true, pdfstartview=XYZ, plainpages=false}

\makeatletter
\let\oldforeign@language\foreign@language
\DeclareRobustCommand{\foreign@language}[1]{%
  \lowercase{\oldforeign@language{#1}}}
\theoremstyle{plain}

\theoremstyle{remark}

\makeatother

\definecolor{blue1}{RGB}{33,92,175}

\newcommand{\bfg}[1]{\boldsymbol{#1}}
\newcommand{\bfb}[1]{\boldsymbol{\rm #1}}

\newcommand{\nx}{n}
\newcommand{\nk}{k}

\usepackage{float}

\acrodef{avr}[AVR]{automatic voltage regulator}
\acrodef{pss}[PSS]{power system stabilizer}
\acrodef{ode}[ODE]{ordinary differential equation}
\acrodef{dae}[DAE]{differential algebraic equation}
\acrodef{lte}[TE]{truncation error}
\acrodef{sg}[SG]{synchronous generator}
\acrodef{tds}[TDS]{time-domain simulation}
\acrodef{tm}[TM]{Trapezoidal Method}
\acrodef{fem}[FEM]{forward Euler method}
\acrodef{bem}[BEM]{backward Euler method}
\acrodef{tg}[TG]{turbine governor}
\acrodef{der}[DER]{distributed energy resource}
\acrodef{pll}[PLL]{phase-locked loop} 
\acrodef{qss}[QSS]{quantized state system} 
\acrodef{qbi}[QBI]{Quantization Based Integration}
\acrodef{tdi}[TDI]{Time Domain Integration}
\acrodef{qss1}[QSS1]{first-order quantized state system}
\acrodef{qss2}[QSS2]{second-order quantized state system}
\acrodef{qss3}[QSS3]{third-order quantized state system}
\acrodef{i}[I]{Integral}
\acrodef{pi}[PI]{Proportional-Integral}
\acrodef{p}[P]{Proportional}
\acrodef{liqss}[LIQSS]{linearly implicit QSS}
\acrodef{ab}[AB]{Adams–Bashforth}
\acrodef{aiits}[AIITS]{all-island Irish transmission system}
\acrodef{devs}[DEVS]{discrete event system specification} 
\acrodef{qe}[QE]{quantization event}

\title{On the Duality Between Quantized Time and States in Dynamic Simulation}

\author{Liya Huang and Georgios Tzounas,~\IEEEmembership{IEEE Member}
\thanks{The authors are with the School of Electrical and Electronic Engineering, University College Dublin, Dublin, D04V1W8, Ireland. \\
Corresponding author's e-mail: \protect\href{http://georgios.tzounas@ucd.ie}{georgios.tzounas@ucd.ie}.}
}
    
\begin{document}

\maketitle
\pagestyle{plain} 
\IEEEpeerreviewmaketitle
 
\begin{abstract}
This letter introduces a formal duality between discrete-time and quantized-state numerical methods.  We interpret \ac{qss} methods as integration schemes applied to a dual form of the system model, where time is seen as a state-dependent variable.  This perspective enables the definition of novel \ac{qss}-based schemes inspired by classical time-integration techniques.  As a proof of concept, we illustrate the idea by introducing a QSS Adams–Bashforth method applied to a test equation.  We then move to demonstrate how the proposed approach can achieve notable performance improvements in realistic power system simulations.
\end{abstract}


\section{Introduction}
\label{sec:intro}

The numerical simulation of power system dynamics inherently requires discretization. Classical Runge–Kutta and multistep methods are based on time discretization, where the system dynamics are approximated at a sequence of discrete time points.  An alternative paradigm is offered by the family of \acf{qss} methods~\cite{kofman2001quantized,kofman2002second,kofman2006third}, where the core idea is to quantize the state variables instead of discretizing time.  These methods can offer computational advantages in problems with localized activity, where some states require more frequent updates than others.  This letter introduces a new perspective that exploits classical integration techniques to guide the design of novel and useful \ac{qss}-based schemes. 

The family of \ac{qss} methods was systematically introduced by Kofman within the \ac{devs} framework \cite{kofman2001quantized}.  The development began with the \ac{qss1} method, and was subsequently extended to higher-order variants designed to improve accuracy~\cite{kofman2002second,kofman2006third}.  Implicit versions were later developed to enhance numerical stability in stiff systems; see, e.g.,~\cite{di2019improving}.
The application of \ac{qss} methods to power systems has so far been limited. Existing studies have primarily considered the simulation of single converters and synchronous machines connected to an infinite bus \cite{di2019improving, gholizadeh2023evaluation}.
A key reason behind this limited application is the asynchronous nature of \ac{qss} methods, where, in principle, each state variable updates at its own rate. This structure complicates their use in large, coupled systems that require coordinating state updates.

In this letter, we revisit \ac{qss} methods through a new lens, drawing analogies with classical time-integration schemes, and show how this leads to new \ac{qss}-based approaches that can be used in conjunction with standard power system solvers.

\section{Duality of State and Time Discretization}
\label{sec:dual}

The central idea behind \ac{qss} is to approximate the solution of differential equations by discretizing the states rather than time.  The simplest and most intuitive variant is the first-order scheme, \ac{qss1}.  Consider the scalar equation:
\begin{equation}
\label{eq:ode}
x'(t) = f(x(t)) \ , \quad x \in \mathbb{R}
\end{equation}
\ac{qss1} tracks $x(t)$ and updates a piecewise-constant signal $q(t)$ whenever $x(t)$ changes by a fixed amount $\Delta q$, called the \textit{quantum}. 
The signal $q(t)$ acts as a quantized approximation of $x(t)$, initialized as $q(0) = x(0)$, and defined by \cite{kofman2001quantized}:
\begin{equation*}
q(t) = 
\begin{cases} 
x(t)  \ \ \ \ \  \text{if } |x(t)- q(t^-)| \geq \Delta q \\ 
q(t^-) \ \ \; \,  \text{otherwise}
\end{cases}
\end{equation*}

At any time $t$, the state derivative is evaluated using $q(t)$.  That is, \eqref{eq:ode} is approximated as:
\begin{equation}
\label{eq:qs}
x'(t) = f(q(t)) 
\end{equation}
Since $q(t)$ changes only when 
$x(t)$ deviates from it by more than $\Delta q$, the right-hand side remains constant in-between, i.e., between \textit{\acp{qe}}. Thus, in \ac{qss1}, $x(t)$ is piecewise linear between successive updates of $q(t)$.
Let $t_\nk$ denote the time of the 
$\nk$-th \ac{qe}. At this time, we have that $q(t_\nk)=x(t_\nk)$.
Then, for $t \in [t_{\nk}, t_{\nk+1})$, the state evolves as:
\begin{equation}
 x(t) = x(t_\nk) + (t-t_\nk)  f(x(t_\nk)) \ ,  \quad t \in [t_\nk, t_{\nk+1})
\label{eq:xt}
\end{equation} 
The next \ac{qe} occurs at $t=t_{\nk+1}$, when, by definition, the following condition is satisfied:
\begin{equation}
|x(t) - x(t_\nk)| = \Delta q  
\label{eq:qe}
\end{equation}
Substituting \eqref{eq:xt} in \eqref{eq:qe} at $t=t_{\nk+1}$, we get
%
$\Delta t_{\nk} | f(x (t_\nk))| = \Delta q $,
%
where $\Delta t_{\nk} = t_{\nk+1}- t_\nk$. 
The integration time step is:
\begin{equation}
t_{\nk+1} = t_{\nk} + \Delta q /{ |f(x(t_\nk))|}
\label{eq:qss:dt}
\end{equation}

Equation~\eqref{eq:qss:dt} reveals a formal analogy, specifically, it is equivalent to forward Euler method applied to the following \textit{dual system} where time evolves as a state-dependent variable:
\begin{equation}
\label{eq:tdot}
 t' (x) = \phi  (t (x))  
\end{equation}
with $\phi = 1 / |f|$.  That is,
%
$t_{\nk+1} = t_{\nk} + \Delta q \; \phi(t_\nk)$.
%
This dual perspective allows us to 
interpret \ac{qss} methods as discrete approximations of \eqref{eq:tdot}.
The following implications are relevant:  
(i) if $x'=0$, then $t'\rightarrow \infty$.  In other words, \ac{qss} jumps directly to the final simulation time if the system has reached steady state, skipping unnecessary computations.  
(ii) if $x' \rightarrow \infty$, then $t'\rightarrow 0$.  In other words, \ac{qss} becomes inefficient in the presence of high-frequency noise or instability.  A trade-off between efficiency and temporal accuracy can be achieved by enforcing a suitable minimum time increment.

The duality above offers valuable insight into the numerical behavior of \ac{qss} methods.  In particular, since~\eqref{eq:qss:dt} mirrors forward Euler applied to \eqref{eq:tdot}, it follows that \ac{qss1} inherits its numerical instability.  In this case, however, the instability does not cause unbounded growth of the state error, but rather \textit{destabilizes the \ac{qe} timing}: when $\Delta q$ is too large, computed event times may deviate significantly from the point at which the state actually crosses the quantization thresholds.  This motivates the design of novel higher-order  methods.  In this letter, we introduce a second-order Adams–Bashforth variant:
\begin{equation}
t_{\nk+1} =  t_{\nk} + \Delta q \; [
1.5\phi (t_\nk) - 0.5  \phi (t_{\nk-1}) 
]
\label{eq:2ab}
\end{equation}  
Note that \eqref{eq:2ab} improves the derivative approximation and thus the timing of \acp{qe}.  Thus, it is conceptually different from, e.g., \cite{kofman2002second}, which focuses on state reconstruction by making $q(t)$ piecewise linear. These two enhancements are independent and can be conveniently combined.  For simplicity, in this letter
we retain the piecewise-constant (QSS1) structure of~$q(t)$.  
\vspace{-2.5mm}
\begin{figure}[ht!]
    \centering
    \begin{subfigure}{0.48\columnwidth}
  \includegraphics[width=\linewidth]{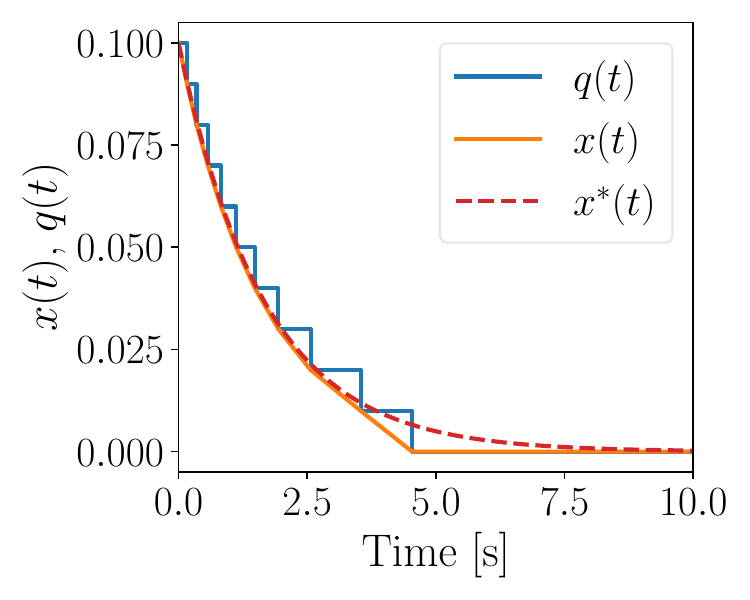}
        \caption{$\Delta q =0.01$.}
        \label{fig:sub1}
    \end{subfigure}
    \hfill
    \begin{subfigure}{0.48\columnwidth}
\includegraphics[width=\linewidth]{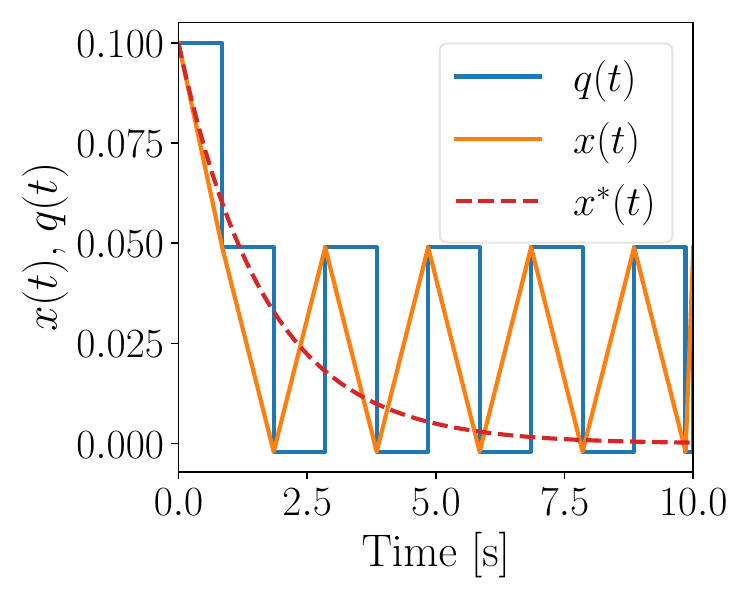}
        \caption{$\Delta q =0.051$.}
        \label{fig:sub2}
    \end{subfigure}
    \vspace{-1.5mm} 
    \caption{Test equation $x'=-0.6 x$: QSS1 solution.}
 \label{fig:qss_comparison}
\end{figure}

\vspace{-2mm}

\begin{figure}[ht!]
    \centering
    \begin{subfigure}{0.48\columnwidth}
  \includegraphics[width=\linewidth]{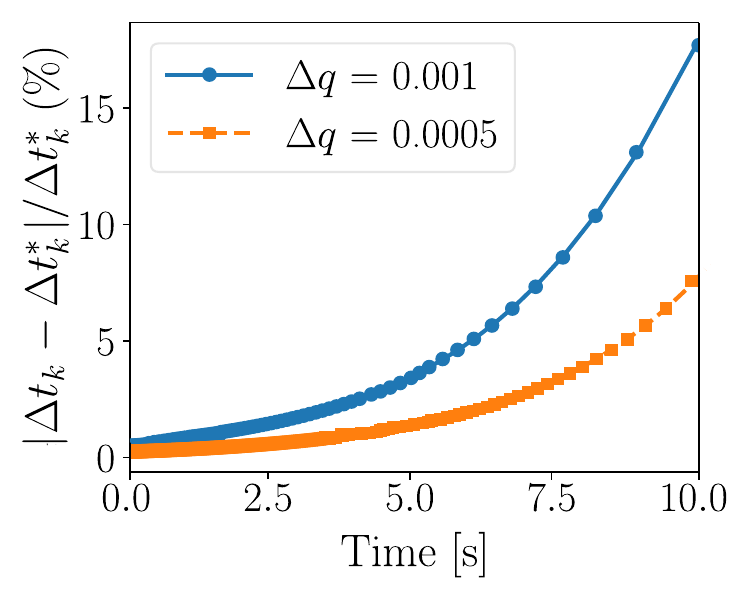}
        \caption{QSS1.}
        \label{fig:sub3}
    \end{subfigure}
    \hfill
    \begin{subfigure}{0.48\columnwidth}
   \includegraphics[width=\linewidth]{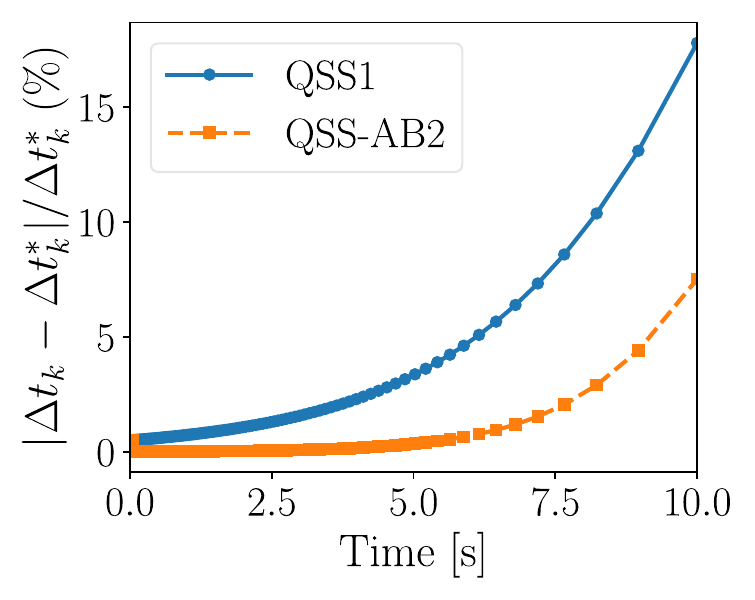}
        \caption{ $\Delta q = 0.001$.}
        \label{fig:sub4}
    \end{subfigure}
    \vspace{-2.5mm} 
    \caption{Test equation $x'=-0.6 x$: Event quantization timing errors.}
    \label{fig:qss_error}
\end{figure}

As an illustrative example, Fig.~\ref{fig:qss_comparison} shows the numerical solution of the test equation $x'(t) = -0.6x(t)$, with $x(0)=0.1$, obtained using \ac{qss1}.  For reference, the exact analytical solution $x^*(t)$ is also included.  As expected, increasing $\Delta q$ leads to larger numerical errors.  For sufficiently large $\Delta q$ (in this case for $\Delta q>0.05$), \ac{qss1} is no longer able to track the exponential state decay. 
The effect of the quantum size $\Delta q$ on the \ac{qe} timing accuracy of \ac{qss1} is illustrated in Fig.~\ref{fig:sub3}.  The relative error 
$|\Delta t_\nk - \Delta t_\nk^*|/\Delta t_\nk^*$ quantifies the deviation between the computed time step $\Delta t_\nk$ and the exact interval $\Delta t^*_\nk$ at which $x(t)$ crosses the quantization thresholds.  As shown in Fig.~\ref{fig:sub4}, QSS-AB2 significantly reduces this error, providing improved event timing accuracy under the same $\Delta q$.

\vspace{-2mm}

\vspace{-4mm}

\section{QSS-Based Time Step Control}
\label{sec:control}

In the preceding analysis we used a scalar equation to illustrate the key idea of viewing \ac{qss} as a time-discretization applied to a dual system, where the roles of time and state are interchanged.  We now generalize this perspective to systems of equations, focusing on dynamical models that describe the behavior of power systems.  These models can be expressed in the form of explicit \acp{dae}:
\begin{equation}
 \begin{bmatrix}
\bfb I_\nx &  \bfb 0_{\nx\times m} \\
\bfb 0_{m\times \nx} & \bfb 0_{m} 
\end{bmatrix}
\begin{bmatrix}
\bfg x'(t) \\
\bfg y'(t)
\end{bmatrix}
= 
\begin{bmatrix}
    \bfg f(\bfg x(t),\bfg y(t)) \\
    \bfg g(\bfg x(t),\bfg y(t)) 
\end{bmatrix}
\label{eq:dae}
\end{equation}
where $\bfg{x}(t) \in \mathbb{R}^\nx$ and $\bfg{y}(t)\in \mathbb{R}^m$ are the state and algebraic variables, respectively. The functions $\bfg{f}$ and $\bfg{g}$ define the differential and algebraic equations, while $\bfb I_\nx$ is the $\nx \times \nx$ identity matrix and $\bfb{0}_{\nx \times m}$ the $\nx \times m$ zero matrix.
We associate a local dual system to each differential equation, as follows:
\begin{equation}
 t_i' (\bfg x, \bfg y) = \phi_i (t_i (\bfg x, \bfg y)) 
\label{eq:t}
\end{equation}
where $\phi_i = 1/|f_i(\bfg x, \bfg y)|$ for $i \in \{1, \dots, \nx\}$.  By applying \eqref{eq:2ab}, we obtain $\nx$ candidate local time steps:
\begin{equation}
\Delta t_{\nk,i} =  \Delta q 
[ 1.5\phi_i (t_{\nk,i}) - 0.5 \phi_i (t_{{\nk-1},i}) ]
\label{eq:2abp}
\end{equation}  

Implementing such asynchronous system updates raises two issues.  First, algebraic equations do not have finite, nonzero time constants, preventing the application of QSS-based stepping to the full \ac{dae} model.  Second, each \ac{qe} may affect multiple variables whose dynamics are coupled, but the causal ordering of their updates is not uniquely defined and thus depends on heuristics.  

In this letter we address these issues by imposing a global time step for the entire system.  In particular, we compute $\Delta t_{\nk,i} $ for each equation from \eqref{eq:2abp} and select:
\begin{equation}
\label{eq:qss:min:dtki}
 \Delta t_{\nk} = \min\{\Delta t_{\nk,i}\} \ , \quad i \in \{1, \dots, \nx \}
\end{equation}

This allows combining novel QSS-based schemes like \eqref{eq:2abp} with standard solvers such as the \ac{tm}.

We note that in this formulation, \acp{qe} are not enforced; instead, $\Delta q$ acts as a parameter that governs time step-size control.  This ensures that no state overshoots its quantization threshold, but can lead to conservative results, as all equations are synchronized with the most sensitive one.  To reduce conservativeness and improve efficiency, we allow the quantum size $\Delta q$ to vary dynamically during the simulation, thereby indirectly controlling the timing of \acp{qe}.  Specifically, we propose updating the $\Delta q$ using a PI control rule, directly inspired by classical step-size adaptation strategies based on local error estimates \cite{soderlind2002automatic,huang2024pi}:
\begin{equation}
  \Delta q_{\nk+1} = \left( {\rm tol}/{|\sigma_\nk|}\right)^{\alpha} \left({|\sigma_{\nk-1}|}/{|\sigma_\nk|}\right)^{\beta} \Delta q_\nk
\label{eq:qss:pi}
\end{equation}
where $\sigma_\nk$ is the estimated event timing error at $t_\nk$; $\rm tol$ is a prescribed error tolerance; and $\alpha$, $\beta$ are control parameters.  

The combination of \eqref{eq:2abp}–\eqref{eq:qss:pi}, that is, time-step selection based on duality-inspired QSS schemes, standard implicit solvers, and adaptive quantum size control, defines a novel framework that we further explore in the next section.

\section{Case Study}
\label{sec:case}

This section presents simulation results based on a dynamic model of the \acf{aiits} \cite{tzounas2020theory}.  The model consists of 1503 buses, 1,851 lines and transformers, 22 synchronous machines, 169 wind power plants and 245 loads.  Simulations are carried out with Dome \cite{milano2013python}, on a computer equipped with an Intel Xeon E3-1245 v5 processor, 16~GB of RAM, and a 64-bit Linux OS.

Two disturbances are considered: 
1) a three-phase fault at bus~1238 occurring at $t=1$~s and cleared by tripping a line connected to the faulted bus after 80~ms; 
2) a loss of 142.06~MW load connected to buses 91-102. All tests are carried out using \ac{tm} as the underlying integration method, considering a simulation duration of 20~s.
For each disturbance, the following configurations are evaluated:  
i) fixed time step; 
ii) QSS1-Sync, where $\Delta t =  \min \{ \Delta q \phi_i (t_{\nk,i}) \}$, $i \in \{1, \dots, \nx \}$, with constant $\Delta q$; 
iii) QSS-AB2, where $\Delta t_\nk$ is defined through \eqref{eq:2abp}, \eqref{eq:qss:min:dtki}, with constant $\Delta q$; 
iv) QSS-AB2-Ad, where $\Delta t_\nk$ is defined through \eqref{eq:2abp}, \eqref{eq:qss:min:dtki}, with adaptive $\Delta q$ updated via the PI control rule \eqref{eq:qss:pi}, using parameters $\alpha = 0.5$, $\beta = 0$, ${\rm tol} = 0.2\times 10^{-1}$.  For QSS-AB2-Ad, the initial $\Delta q$ is set to 0.2 and a maximum value of 4 is imposed to avoid excessive time steps. The error $\sigma_\nk$ is taken as the difference between $\Delta t_{k}$ computed with $\Delta q_k$ and $\Delta q_k/2$, similar to embedded error estimates used in variable-step methods.

\begin{table}[ht!]
\centering
\caption{Three-phase fault: Performance comparison.}  
\label{tab:fault}
\setlength{\tabcolsep}{3pt}  
\vspace{-1mm}
\begin{tabular}{@{}llrr@{}}  
\toprule  
Time step & Quantum  & Time~[s] & Avg. error [$\times 10^{-5}$] 
\\
\midrule 
Fixed [$\Delta t=0.001$~s] & -  & $32.2$  & -     \\
Fixed [$\Delta t=0.01$~s] & - & $5.8$  & $ 42.92 $ 
\\
QSS1-Sync      & Fixed [$\Delta q=0.01$]  &  $9.2$ & $  42.76$  \\
QSS1-Sync      & Fixed [$\Delta q=0.24$] &  $5.6$ & $71.89 $  \\
QSS-AB2   & Fixed [$\Delta q=0.24$]   &  $4.6$ & 1.29
$ $
\\
QSS-AB2-Ad    & PI control  & $4.1$  & 
$0.14$
\\
\bottomrule  
\end{tabular}
\end{table}

\vspace{-3mm}

\begin{table}[ht!]
\centering
\caption{Load loss: Performance comparison.}  
\label{tab:load}
\setlength{\tabcolsep}{3pt}  
\vspace{-1mm}
\begin{tabular}{@{}llrr@{}}  
\toprule  
Time step & Quantum & Time~[s] & Avg. error [$\times 10^{-5}$]
\\
\midrule 
Fixed [$\Delta t=0.001$~s] & -  & $47.5$  & -     \\
Fixed [$\Delta t=0.01$~s] & - &   $6.3$ & $ 153.46$ 
\\
QSS1      & Fixed [$\Delta q=0.01$]  &  $10.0$ & $ 153.81  $  \\
QSS1      & Fixed [$\Delta q=0.24$] &  $ 5.6$ & $189.77 $  \\
QSS-AB2   & Fixed [$\Delta q=0.24$]   & $5.1$  & $ 8.34 $  \\
QSS-AB2-Ad    & PI control  & $4.3$   & 
$0.06$
\\
\bottomrule  
\end{tabular}
\end{table}

\vspace{-3mm}

\begin{figure}[ht!]
    \centering
\vspace{-3mm}\includegraphics[width=0.82\linewidth]
  {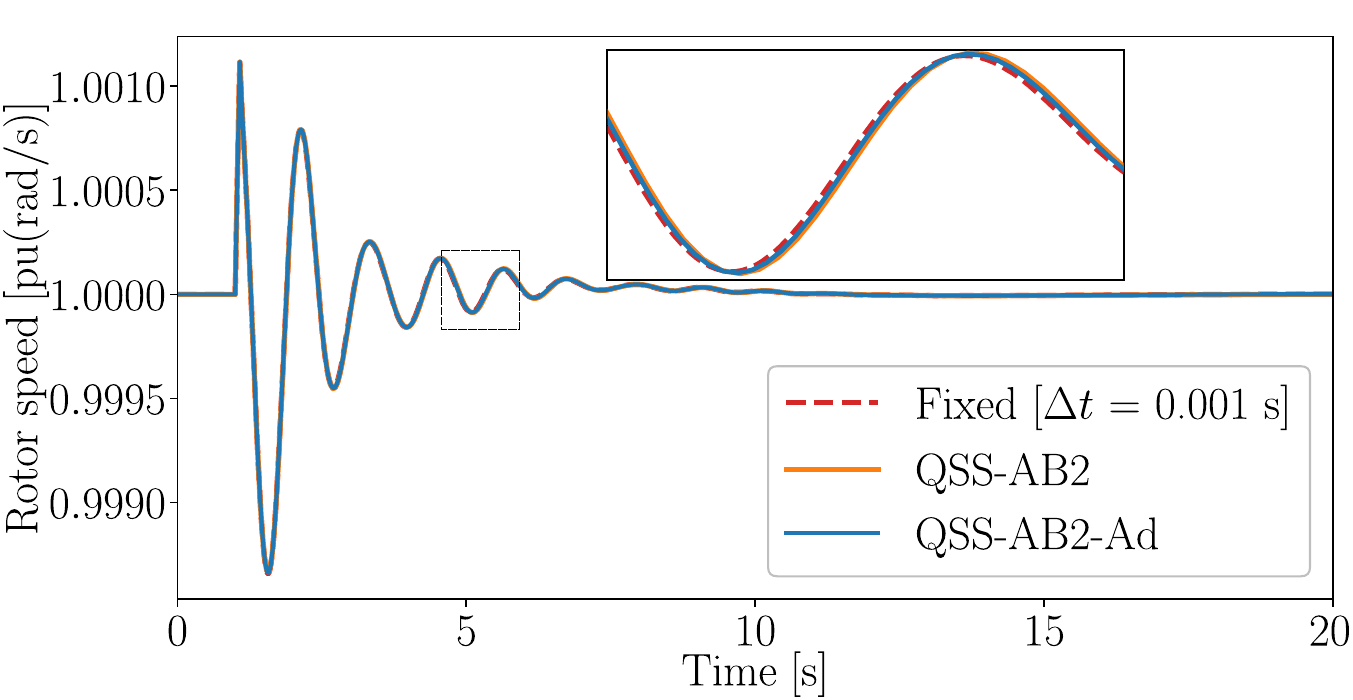}
   \vspace{-1mm}
    \caption{Three-phase fault: Rotor speed of machine at bus~1237.}
    \label{fig:fault}
\end{figure}

Tables~\ref{tab:fault} and \ref{tab:load} report the performance of the examined configurations under the examined disturbances.  In both cases, the
numerical error is measured for the rotor speed of a representative machine (at bus~1237), using \ac{tm} with fixed step size $0.001$~s as the reference.  Increasing the step to $0.01$~s yields faster simulations and a moderate error increase.  Using QSS1-Sync with $\Delta q = 0.01$ exhibits similar accuracy to \ac{tm} with $\Delta t = 0.01$~s but results in higher runtime.  Increasing $\Delta q$ to $0.24$ improves runtime at the expense of accuracy.  In contrast, QSS-AB2 with the same $\Delta q = 0.24$ achieves lower runtime and much lower error compared to \ac{tm} with $\Delta t = 0.01$~s. 
Finally, QSS-AB2-Ad dynamically adjusts $\Delta q$ to further improve both accuracy and runtime.
  
\begin{figure}[ht!]
\vspace{-3mm}
\centering
\includegraphics[width=0.82\linewidth]{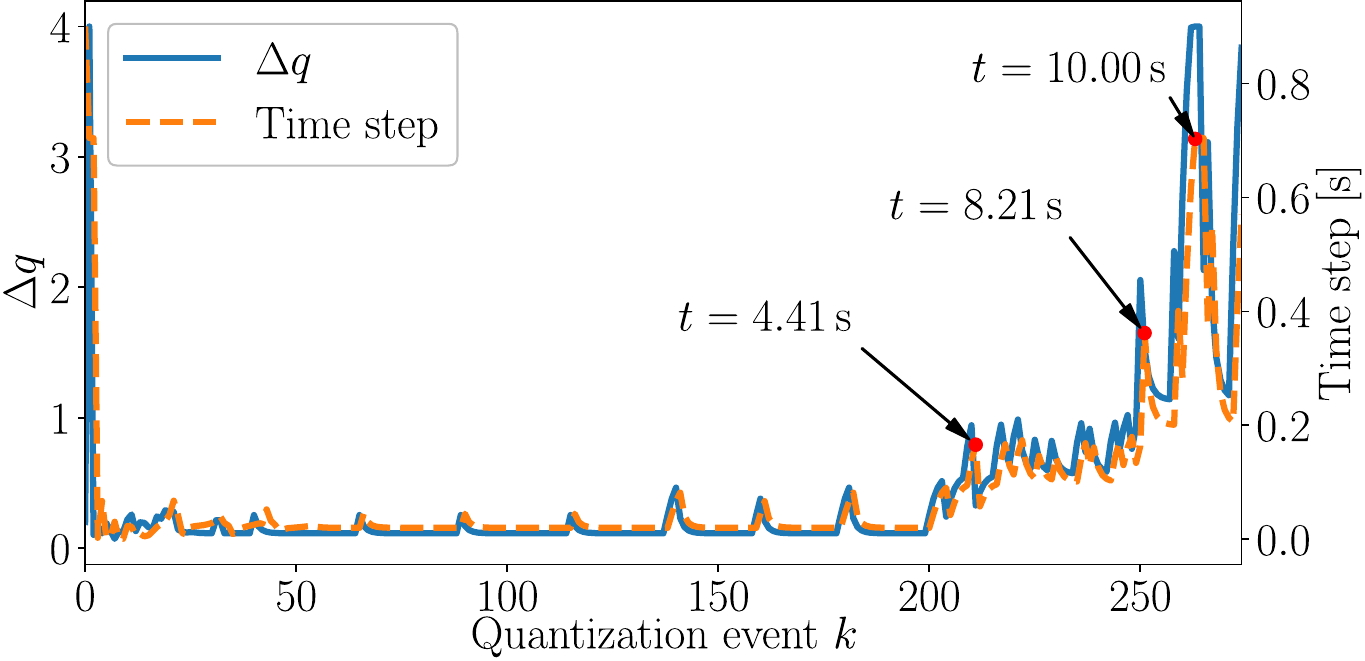}  
\vspace{-1mm}
\caption{Three-phase fault: Variations of $\Delta q$ and $\Delta t$, QSS-AB2-Ad.}
\label{fig:q}
\end{figure}

The rotor speed of the machine at bus 1237 following the fault is shown in Fig.~\ref{fig:fault}.  Both QSS-AB2 and QSS-AB2-Ad accurately track the reference solution, consistent with the errors reported in Table~\ref{tab:fault}.  For the same disturbance, Fig.~\ref{fig:q} further illustrates the behavior of QSS-AB2-Ad, showing how $\Delta q$ and $\Delta t$ evolve over the course of \acp{qe} during the simulation.  In particular, both remain small during the transient and increase significantly as the system approaches steady state.

\vspace{-1mm}

\section{Conclusion}

This letter introduces a novel perspective on quantized-state simulation by formalizing a duality between time and state discretization.  To illustrate  this perspective, we define a QSS-based, second-order Adams–Bashforth method and describe how it can be combined with standard implicit power system solvers to achieve significant speedups while maintaining accuracy.  Further performance gains through adaptive quantum control are also discussed.  The proposed approach opens directions for a new family of numerical methods that integrate time discretization and state quantization.

\vspace{-1mm}

\bibliographystyle{IEEEtran}
\bibliography{references}

\end{document}